\documentclass[10pt,twocolumn]{article}
\usepackage[margin=0.75in]{geometry}
\usepackage[T1]{fontenc}
\usepackage[utf8]{inputenc}
\usepackage{newtxtext,newtxmath}
\usepackage{microtype}
\usepackage{graphicx}
\usepackage{booktabs}
\usepackage{multirow}
\usepackage{amsmath}
\usepackage{array}
\usepackage{enumitem}
\usepackage{caption}
\usepackage{subcaption}
\usepackage{xcolor}
\usepackage{hyperref}
\usepackage{algorithm}
\usepackage{algpseudocode}
\usepackage[numbers,sort&compress]{natbib}
\usepackage{dblfloatfix}
\hypersetup{colorlinks=true,citecolor=blue,linkcolor=blue,urlcolor=blue}
\setlist[itemize]{leftmargin=*,itemsep=2pt,topsep=2pt}
\setlist[enumerate]{leftmargin=*,itemsep=2pt,topsep=2pt}
\title{A Time-Consistent Benchmark for Repository-Level Software Engineering Evaluation}

\author{Xianpeng (Simon) Sun, Haonan Sun, Tian Yu, Sheng Ma, Qincheng Zhang, Lifei Rao, Chen Tian\thanks{This work was conducted independently in the author's personal capacity. The views expressed in this paper are those of the author and do not necessarily reflect the views of Microsoft.}\\
\texttt{sisun@microsoft.com}}

\date{}

\begin{document}
\maketitle

\begin{abstract}
Evaluation of repository-aware software engineering systems is often confounded by synthetic task design, prompt leakage, and temporal contamination between repository knowledge and future code changes. We present a time-consistent benchmark methodology that snapshots a repository at time $T_0$, constructs repository-derived code knowledge using only artifacts available before $T_0$, and evaluates on engineering tasks derived from pull requests merged in the future interval $(T_0, T_1]$. Each historical pull request is transformed into a natural-language task through an LLM-assisted prompt-generation pipeline, and the benchmark is formalized as a matched A/B comparison in which the same software engineering agent is evaluated with and without repository-derived code knowledge while all other variables are held constant. We also report a baseline characterization study on two open-source repositories, DragonFly and React, using three Claude-family models and four prompt granularities. Across both repositories, file-level F1 increases monotonically from minimal to guided prompts, reaching 0.8081 on DragonFly and 0.8078 on React for the strongest tested model. These results show that prompt construction is a first-order benchmark variable rather than a packaging detail. The benchmark protocol is therefore designed to separate realism, leakage control, and knowledge-augmentation effects in a way that supports stable and repeatable comparisons across systems.
\end{abstract}

\section{Introduction}
Repository-aware coding agents increasingly rely on external knowledge layers: retrieval systems, code graphs, semantic indices, repository memories, and workflow traces built over a codebase. Yet evaluation practice still tends to collapse multiple factors into a single score. A benchmark may unintentionally reward prompt verbosity, future-information leakage, or hidden implementation hints rather than genuine repository understanding.

This problem is especially acute for \emph{repository-knowledge-augmented systems}, whose purpose is not merely to answer a task prompt, but to build and exploit structured knowledge about a repository. A valid benchmark for such systems must satisfy three constraints simultaneously. First, tasks should reflect authentic software engineering work rather than synthetic exercises. Second, the system should only have access to information that would have been available at the time the task arose. Third, the contribution of repository-derived code knowledge should be isolated from confounders such as changes in prompt detail, model configuration, or execution environment. Existing repository-level benchmarks have already made important progress on realism, task diversity, execution-based validation, benchmark freshness, and contamination resistance; our focus is narrower and more causal. We seek to estimate the \emph{marginal effect of repository-derived code knowledge itself} under fixed repository snapshot, fixed task, fixed model, fixed execution environment, and fixed evaluation metric.

We address these requirements with a \textbf{time-consistent benchmark protocol} based on historical pull requests. A repository is snapshotted at time $T_0$; repository knowledge is constructed only from artifacts available at that snapshot; and evaluation tasks are derived from pull requests merged later in $(T_0,T_1]$. The protocol is designed for \textbf{matched A/B evaluation}: the same SWE agent is run on the same repository snapshot and the same task set under two conditions, once without repository-derived code knowledge and once with it.

In addition to formalizing the benchmark, we report a \emph{baseline-only characterization study}. This study does not yet claim gains from repository-knowledge augmentation. Instead, it quantifies how much benchmark scores move due only to prompt granularity and model capability. That calibration is important because any future knowledge-augmented-vs-baseline delta must be interpreted against this background variance.

\paragraph{Contributions.}
\begin{enumerate}
    \item We define a \textbf{time-consistent benchmark methodology} for repository-aware software engineering systems using historical pull requests and strict pre/post-snapshot separation.
    \item We formalize a \textbf{matched A/B evaluation design} that isolates the causal contribution of repository-derived code knowledge from model, prompt, environment, and task effects.
    \item We integrate a \textbf{prompt-generation design space} into benchmark construction and show empirically that prompt granularity strongly affects measured performance.
    \item We provide a \textbf{baseline characterization study} on DragonFly and React across three models and four prompt granularities, establishing a calibration point for future knowledge-augmented evaluations.
\end{enumerate}

\section{Related Work}
\textbf{Real-world software engineering benchmarks.} SWE-bench established a widely adopted benchmark in which models resolve real GitHub issues from historical repositories under repository snapshots and execution-based verification \citep{jimenez2024swebench}. RepoBench expanded evaluation to repository-level retrieval, completion, and pipeline tasks, highlighting the importance of cross-file context in realistic code assistance \citep{liu2024repobench}. More recent benchmark efforts have broadened language coverage and task diversity, including repository-level execution-based evaluation in SWE-PolyBench and feature-implementation tasks in FEA-Bench \citep{buccholz2025swepolybench, jia2025feabench}. These benchmarks are highly relevant, but they are primarily designed to evaluate overall task performance rather than to isolate the incremental effect of an external repository-knowledge layer under strict pre/post-snapshot separation.

\textbf{Dynamic and contamination-aware benchmark construction.} A second line of work focuses on scalable benchmark generation, historically faithful environment setup, and contamination resistance. Automated benchmark generation pipelines have shown that historically accurate setup and large-scale collection are feasible, but operationally delicate \citep{vergopoulos2025automated}. SWE-MERA emphasizes dynamic refresh, automated task collection, and regular benchmark updates, while SWE-rebench explicitly frames benchmark freshness as a response to contamination risk and uses a continuous pipeline to build decontaminated evaluation sets \citep{swemera2025, badertdinov2025swerebench}. Our proposal is aligned with this direction, but it targets a different question: not merely how to build fresher benchmarks, but how to estimate the causal contribution of repository-derived code knowledge under matched historical conditions.

\textbf{Prompt sensitivity and benchmark validity.} Benchmark scores are often strongly affected by prompt framing, task exposure, and hidden implementation hints. In repository-grounded settings, prompt design is especially consequential because a task prompt can either preserve the need for repository exploration or leak the solution structure. We therefore treat prompt generation as part of benchmark methodology rather than benchmark packaging.

\section{Benchmark Methodology}
A challenge in evaluating repository-aware software engineering systems is to design a benchmark that is both realistic and leakage-resistant. Benchmarks based on synthetic tasks or manually written problem statements often fail to capture the structure of real engineering work. At the same time, naively evaluating on historical repository changes can introduce information leakage if the system has access to code that was added after the task was created.

To address this problem, we design a time-consistent benchmark methodology based on historical pull requests. The key idea is to construct repository-derived code knowledge from a repository snapshot taken at time $T_0$, and then evaluate the system on engineering tasks derived from pull requests merged in the future interval $(T_0,T_1]$. This ensures that the offline knowledge-construction pipeline can only use information that would have been available at the time of the snapshot, while the evaluation tasks correspond to real software engineering work that occurred afterward.

\begin{figure*}[t]
    \centering
    \includegraphics[width=0.80\textwidth]{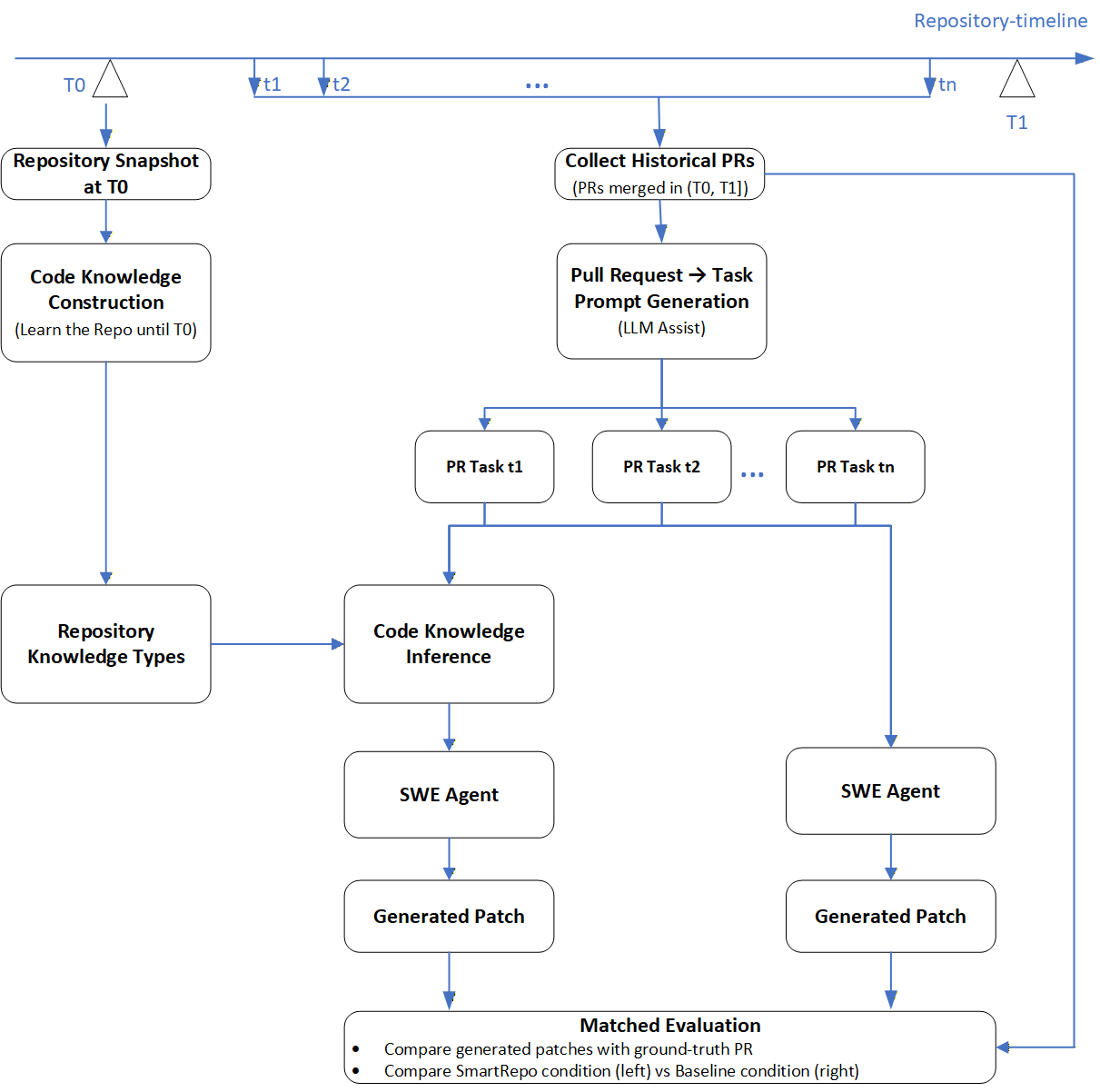}
    \caption{Time-consistent benchmark pipeline. A repository is snapshotted at $T_0$ to construct repository knowledge using only pre-$T_0$ information. Pull requests merged during $(T_0,T_1]$ are converted into natural-language tasks. The same SWE agent is then evaluated in matched conditions with and without repository-derived code knowledge, and outputs are compared against the historical ground-truth pull requests.}
    \label{fig:pipeline}
\end{figure*}

Figure~\ref{fig:pipeline} illustrates the benchmark construction and evaluation flow. Starting from a repository timeline, we first take a snapshot at $T_0$ and build repository-derived code knowledge using only repository state available up to that point. We then collect all pull requests merged during $(T_0,T_1]$ and convert them into natural-language task prompts using an LLM-assisted pipeline. The resulting task set is evaluated under two matched conditions: a baseline SWE agent without repository-derived code knowledge and the same SWE agent augmented with repository-derived code knowledge. Generated solutions are then compared against the corresponding ground-truth pull requests.

This methodology serves two purposes. First, it provides a realistic evaluation setting in which benchmark tasks are grounded in authentic repository evolution. Second, it isolates the contribution of repository-derived code knowledge by ensuring that the only difference between the two evaluation conditions is whether the agent has access to repository-derived code knowledge.

\subsection{Prompt Construction via LLM}

To ensure consistency and reproducibility in benchmark task generation, 
we define a structured prompt construction mechanism that transforms 
historical pull requests into task descriptions at different levels 
of abstraction.

Given a pull request $p_i$, we define a prompt generation function:
\[
t_i = g(p_i, \ell)
\]
where $\ell \in \{\text{minimal}, \text{concise}, \text{contextual}, \text{guided}\}$ 
controls the level of detail.

\textbf{Minimal.}
The prompt contains only a high-level description derived from the PR title.

\textbf{Concise.}
The prompt includes the affected component or module, without revealing implementation details.

\textbf{Contextual.}
The prompt incorporates additional context such as functionality, usage scenario, or failure mode.

\textbf{Guided.}
The prompt provides structured guidance on the nature of the fix, 
while avoiding direct exposure of the ground-truth implementation.

This design enables controlled evaluation of how task specification 
granularity affects software engineering performance, while ensuring 
that prompts do not leak solution-critical information.

\subsection{Time-Snapshot and Matched A/B Design}
To isolate the effect of repository-derived code knowledge on software engineering performance, we adopt a matched A/B evaluation design over historical pull requests. For each repository, we first select a historical snapshot at time $T_0$. Let $R_{T_0}$ denote the repository state at that snapshot. An offline knowledge-construction pipeline builds repository-derived code knowledge exclusively from information available in $R_{T_0}$, including source files, commit history, and repository documentation. No information introduced after $T_0$ is allowed to participate in knowledge construction. In particular, post-$T_0$ pull-request descriptions, diffs, review comments, newly introduced files, and downstream test outcomes are outside the admissible knowledge boundary. This explicit pre/post split is intended to separate historically available repository evidence from future task outcomes.

We then collect the set of pull requests merged during the interval $(T_0,T_1]$:
\begin{equation}
P = \{p_i \mid p_i \text{ is merged in } (T_0,T_1]\}.
\end{equation}
Each pull request $p_i\in P$ is transformed into a natural-language software engineering task through a prompt generation function $g(\cdot)$, producing the benchmark task set
\begin{equation}
D = \{t_i = g(p_i) \mid p_i \in P\}.
\end{equation}
Each task $t_i$ therefore corresponds to a real historical code change that occurred after the repository snapshot.

For every task $t_i\in D$, we evaluate the same SWE agent under two matched conditions.
\begin{itemize}
    \item \textbf{Baseline condition.} The agent operates directly on the repository snapshot $R_{T_0}$ without access to repository-derived code knowledge.
    \item \textbf{Knowledge-augmented condition.} The agent operates on the same repository snapshot $R_{T_0}$, but is augmented with repository-derived code knowledge constructed from that snapshot.
\end{itemize}
Let $A_{\text{base}}(t_i)$ denote the solution produced by the baseline agent for task $t_i$, and let $A_{\text{aug}}(t_i)$ denote the solution produced by the knowledge-augmented agent for the same task. All other factors are held constant across the two conditions, including the underlying SWE agent and language model, the repository snapshot $R_{T_0}$, the task prompt $t_i$, the execution environment, and the evaluation metric. As a result, the only experimental variable is whether the agent has access to repository-derived code knowledge. This paired design isolates the marginal contribution of repository-derived code knowledge to repository-grounded task solving.

For a system $S\in\{\text{base},\text{aug}\}$, the overall benchmark score is computed as
\begin{equation}
\mathrm{Score}(S)=\frac{1}{|D|}\sum_{t_i\in D} \mathrm{metric}(A_S(t_i), y_i),
\end{equation}
where $y_i$ denotes the ground-truth outcome associated with the historical pull request for task $t_i$. This is not an online product experiment; it is a controlled benchmark-style matched comparison whose purpose is to estimate the causal impact of repository-derived code knowledge on software engineering performance.

\subsection{Prompt Generation from Historical Pull Requests}
A non-trivial challenge in constructing PR-based benchmarks lies in converting historical pull requests into realistic task prompts. Pull requests typically contain multiple sources of information, including the title, description, code diffs, and discussion comments. If these inputs are converted too aggressively into natural-language prompts, the generated tasks may leak implementation details that would not normally be available in realistic developer requests. On the other hand, if too little information is retained, the resulting tasks may become under-specified and no longer represent meaningful software engineering requests.

During early experiments, we found that the granularity of prompt generation has a substantial impact on benchmark outcomes. In particular, prompt formulations that contain overly specific implementation details can significantly improve baseline performance, thereby reducing the apparent gain from repository-derived code knowledge. Conversely, prompts that are too short often omit essential contextual cues, causing both baseline and knowledge-augmented agents to fail.

To study this effect systematically, we implemented four prompt-generation strategies:
\begin{itemize}
    \item \textbf{Minimal (very concise).} A compressed restatement of the high-level change, often close to the pull request title alone.
    \item \textbf{Concise.} A brief task description that preserves the problem type while providing only limited repository grounding.
    \item \textbf{Contextual.} The primary benchmark setting: it preserves the general problem description and affected component, but avoids exposing exact function names, control flow, or explicit repair instructions.
    \item \textbf{Guided.} A richer prompt that incorporates stronger task framing and partial implementation hints extracted from the pull request.
\end{itemize}
Minimal prompts are concise but often under-specify the task. Guided prompts improve absolute performance for both the baseline and knowledge-augmented conditions, but they can also reduce the relative gain attributable to repository-derived code knowledge by revealing likely components, functions, or fix directions. Contextual prompts typically provide the best realism--solvability trade-off, and unless otherwise noted, they are the most defensible primary setting for future knowledge-augmented comparisons.

\subsection{Worked Example of Benchmark Construction}
Consider a repository snapshot $T_0$ that contains a dataset-loading component. A later pull request introduces a null check in metadata parsing to prevent a crash when metadata is missing. The pull request can be converted into the benchmark task: \emph{Fix a bug in dataset metadata parsing within the DatasetLoader module.}

Under the baseline condition, the agent searches the repository for relevant files based on the prompt. Because the prompt does not explicitly mention the affected function, the agent may explore multiple candidate files. Under the knowledge-augmented condition, the agent receives repository-derived code knowledge indicating that \texttt{DatasetLoader} is responsible for dataset configuration parsing and that the relevant metadata pathway is concentrated in a small set of files. The task is considered successful if the generated solution matches the historical change under the chosen evaluation metric.

\section{Experimental Setup}
We evaluate the baseline arm of the benchmark protocol on two open-source repositories: DragonFly and React. For each repository, a historical snapshot is taken prior to the evaluation window, and historical pull requests merged after the snapshot are converted into benchmark tasks following the methodology of Section~3.

\begin{table}[t]
\centering
\caption{Benchmark dataset and evaluation configuration. The protocol itself supports matched knowledge-augmented-vs-baseline comparison; the present section reports only the baseline arm.}
\label{tab:setup}
\begin{tabular}{p{0.30\columnwidth}p{0.60\columnwidth}}
\toprule
Category & Setting \\
\midrule
Repositories & DragonFly and React \\
Task source & Historical PRs merged in $(T_0,T_1]$ \\
Snapshot policy & Knowledge may only be constructed from repository state available at $T_0$ \\
Models & Claude-Sonnet-4, Claude-Sonnet-4.5, Claude-Ops-4.6 \\
Prompt strategy & Minimal, Concise, Contextual, Guided \\
Current reported arm & Baseline only (no repository-knowledge augmentation in this characterization study) \\
Evaluation metrics & File-level precision, recall, and F1 \\
Unit of comparison & Predicted relevant file set vs. PR-modified file set \\
\bottomrule
\end{tabular}
\end{table}

\subsection{Metric Definition}
The current experiments evaluate \emph{file-level localization quality}. For each task, the AI agent produces a set of predicted files that it considers relevant for implementing the requested change. These predictions are compared against the set of files that were modified in the corresponding historical pull request. Let TP denote files suggested by the agent that were indeed modified, FP denote files suggested by the agent but not modified in the ground-truth change, and FN denote files modified in the ground truth but not suggested by the agent.

We report the standard information-retrieval metrics
\begin{align}
\mathrm{Precision} &= \frac{\mathrm{TP}}{\mathrm{TP}+\mathrm{FP}},\\
\mathrm{Recall} &= \frac{\mathrm{TP}}{\mathrm{TP}+\mathrm{FN}},\\
F_1 &= \frac{2\cdot \mathrm{Precision}\cdot \mathrm{Recall}}{\mathrm{Precision}+\mathrm{Recall}}.
\end{align}
Together, these metrics evaluate both the accuracy and completeness of repository-level file prediction. In future matched A/B studies, the same definitions can be used to compare a knowledge-augmented condition directly against the baseline condition.

\section{Results}
\subsection{DragonFly}
\begin{figure}[t]
    \centering
    \includegraphics[width=\columnwidth]{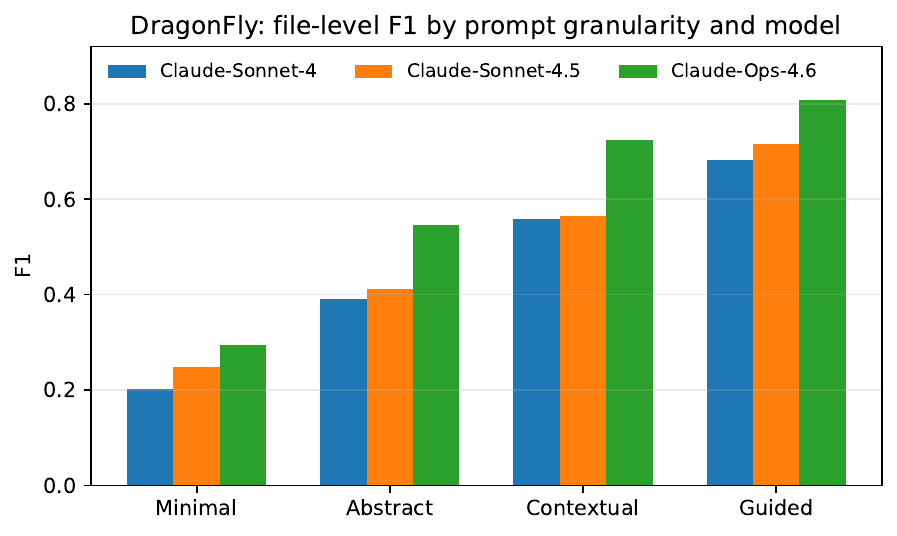}
    \caption{DragonFly F1 by prompt granularity and model.}
    \label{fig:dragonfly-bars}
\end{figure}

\begin{figure}[t]
    \centering
    \includegraphics[width=\columnwidth]{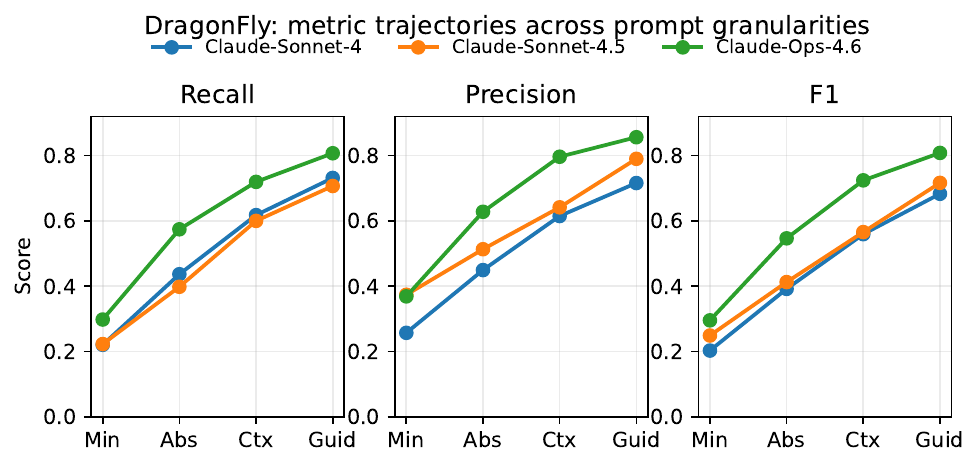}
    \caption{DragonFly metric trajectories across prompt granularities.}
    \label{fig:dragonfly-traj}
\end{figure}

Table~\ref{tab:dragonfly} reports full baseline results on DragonFly. Figure~\ref{fig:dragonfly-bars} summarizes F1 by model and prompt granularity, and Figure~\ref{fig:dragonfly-traj} shows recall, precision, and F1 trajectories across prompt settings.

\begin{table}[t]
\centering
\caption{DragonFly baseline results. All values are file-level scores.}
\label{tab:dragonfly}
\resizebox{\columnwidth}{!}{%
\begin{tabular}{llccc}
\toprule
Model & Prompt & Recall & Precision & F1 \\
\midrule
\multirow{4}{*}{Sonnet-4}
& Minimal & 0.2202 & 0.2569 & 0.2026 \\
& Concise & 0.4365 & 0.4492 & 0.3916 \\
& Contextual & 0.6176 & 0.6148 & 0.5592 \\
& Guided & 0.7314 & 0.7158 & 0.6828 \\
\midrule
\multirow{4}{*}{Sonnet-4.5}
& Minimal & 0.2225 & 0.3736 & 0.2487 \\
& Concise & 0.3980 & 0.5131 & 0.4124 \\
& Contextual & 0.5999 & 0.6414 & 0.5656 \\
& Guided & 0.7068 & 0.7900 & 0.7163 \\
\midrule
\multirow{4}{*}{Ops-4.6}
& Minimal & 0.2977 & 0.3686 & 0.2952 \\
& Concise & 0.5741 & 0.6278 & 0.5464 \\
& Contextual & 0.7194 & 0.7964 & 0.7241 \\
& Guided & 0.8071 & 0.8562 & 0.8081 \\
\bottomrule
\end{tabular}}
\end{table}

Three patterns stand out. First, prompt granularity strongly affects measured performance. On DragonFly, moving from Minimal to Guided increases F1 from 0.2026 to 0.6828 for Sonnet-4, from 0.2487 to 0.7163 for Sonnet-4.5, and from 0.2952 to 0.8081 for Ops-4.6. Second, stronger models consistently improve all three metrics at matched prompt granularity. Third, the largest jump often occurs between Concise and Contextual prompts, suggesting that repository-grounded module information plays a major role in file localization.

\subsection{React}
\begin{figure}[t]
    \centering
    \includegraphics[width=\columnwidth]{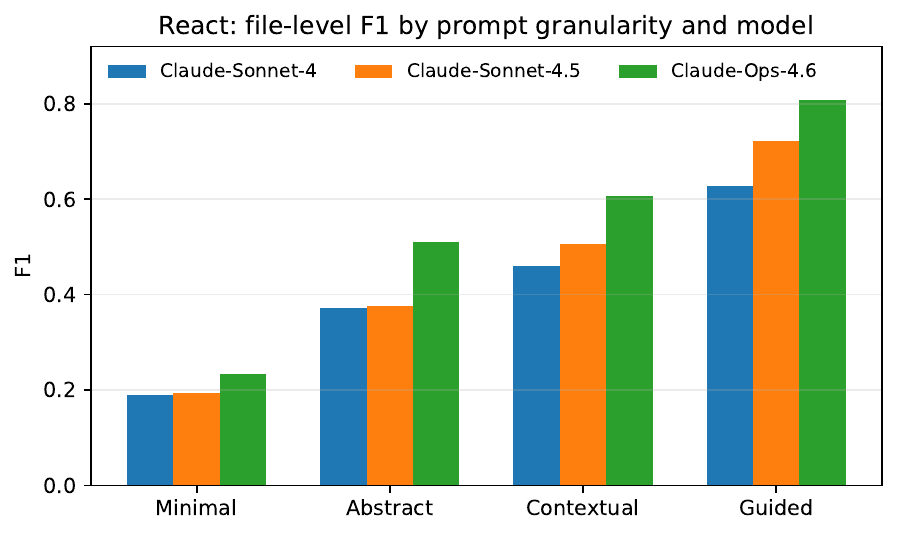}
    \caption{React F1 by prompt granularity and model.}
    \label{fig:react-bars}
\end{figure}

\begin{figure}[t]
    \centering
    \includegraphics[width=\columnwidth]{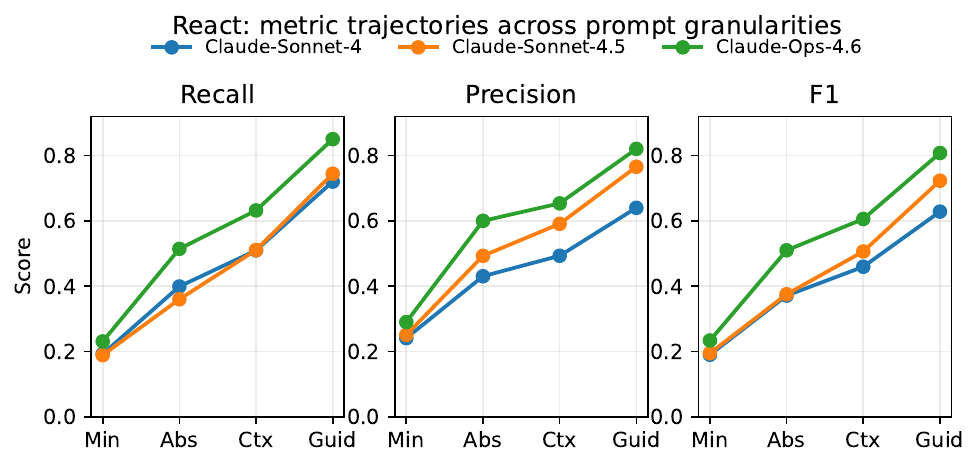}
    \caption{React metric trajectories across prompt granularities.}
    \label{fig:react-traj}
\end{figure}

Table~\ref{tab:react} reports the corresponding baseline results on React. Figures~\ref{fig:react-bars} and \ref{fig:react-traj} visualize the same trend.

\begin{table}[t]
\centering
\caption{React baseline results. All values are file-level scores.}
\label{tab:react}
\resizebox{\columnwidth}{!}{%
\begin{tabular}{llccc}
\toprule
Model & Prompt & Recall & Precision & F1 \\
\midrule
\multirow{4}{*}{Sonnet-4}
& Minimal & 0.1921 & 0.2407 & 0.1898 \\
& Concise & 0.3989 & 0.4303 & 0.3710 \\
& Contextual & 0.5095 & 0.4930 & 0.4594 \\
& Guided & 0.7202 & 0.6399 & 0.6283 \\
\midrule
\multirow{4}{*}{Sonnet-4.5}
& Minimal & 0.1886 & 0.2506 & 0.1946 \\
& Concise & 0.3598 & 0.4928 & 0.3751 \\
& Contextual & 0.5106 & 0.5909 & 0.5062 \\
& Guided & 0.7444 & 0.7657 & 0.7229 \\
\midrule
\multirow{4}{*}{Ops-4.6}
& Minimal & 0.2305 & 0.2899 & 0.2335 \\
& Concise & 0.5142 & 0.6000 & 0.5098 \\
& Contextual & 0.6319 & 0.6536 & 0.6057 \\
& Guided & 0.8503 & 0.8205 & 0.8078 \\
\bottomrule
\end{tabular}}
\end{table}

The React results replicate the DragonFly trend. Under Minimal prompting, file-level F1 remains low for all three models (0.1898--0.2335). Under Guided prompting, F1 reaches 0.6283 for Sonnet-4, 0.7229 for Sonnet-4.5, and 0.8078 for Ops-4.6. This cross-repository consistency is important because it suggests the prompt-sensitivity effect is not an artifact of one repository alone.

\subsection{Task-Level Distributional Analysis}

While the aggregate means in Table~\ref{tab:react} establish the overall trend, they do not reveal whether improvements are broadly distributed across tasks or concentrated in a subset of easy cases. To address this, we analyze task-level outcomes from the React baseline runs.

Each model--prompt condition includes 90 historical PR tasks, enabling us to examine exact-success and exact-failure rates, as well as the full distribution of outcomes, rather than relying solely on averages.

\begin{table}[t]
\centering
\small
\setlength{\tabcolsep}{4pt}
\caption{React task-level baseline summary aggregated across the three models.}
\label{tab:react_task_summary}
\begin{tabular}{lcccc}
\toprule
Prompt & Mean F1 & Prec.=0 & Rec.=1 & Rec.=0 \\
\midrule
Minimal   & 0.242 & 60.2\% & 14.4\% & 60.2\% \\
Concise   & 0.479 & 28.7\% & 31.1\% & 28.7\% \\
Contextual& 0.574 & 18.5\% & 41.5\% & 18.5\% \\
Guided    & 0.770 & 3.5\%  & 59.8\% & 3.5\% \\
\bottomrule
\end{tabular}
\end{table}

Table~\ref{tab:react_task_summary} sharpens the interpretation of the mean curves. The dominant effect of prompt strengthening is not merely a modest upward drift in average F1; it is a large redistribution away from exact failure states. Averaged across models, the zero-precision rate falls from 60.2\% under Minimal prompts to 18.5\% under Contextual prompts and only 3.5\% under Guided prompts, while the perfect-recall rate rises from 14.4\% to 41.5\% and then 59.8\%. In other words, prompt granularity changes the probability that the agent can localize relevant files at all, rather than only improving overlap once localization succeeds.

\begin{figure*}[t]
\centering
\includegraphics[width=0.96\textwidth]{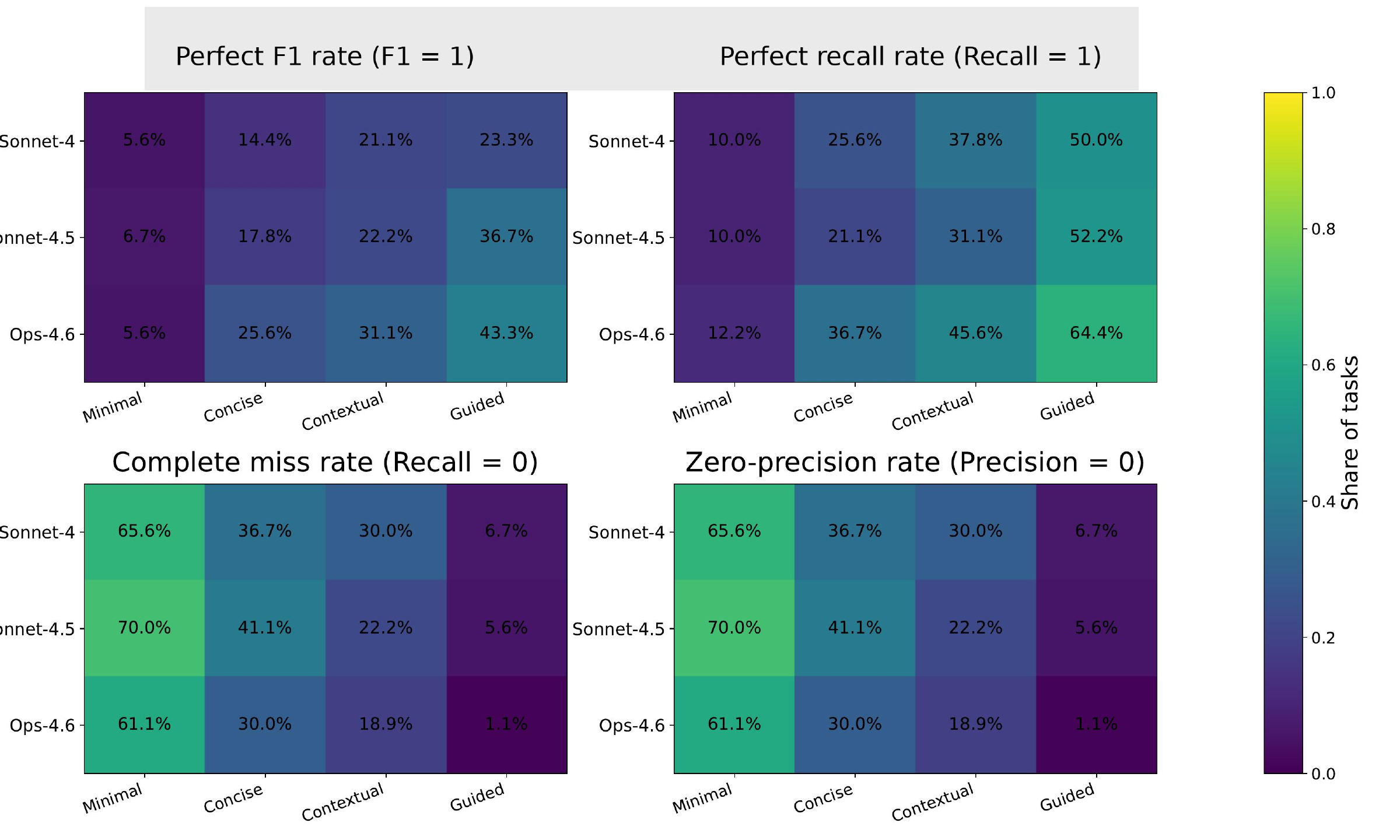}
\caption{React baseline, task-level extreme outcomes. Each cell reports the fraction of historical PR tasks that land in an exact success or exact failure state. The largest movement is from Minimal/Concise to Contextual/Guided: better prompts sharply reduce both complete misses (Recall$=0$) and spurious localization failures (Precision$=0$), while increasing exact localization (Recall$=1$, F1$=1$).}
\label{fig:react-extreme-large}
\end{figure*}

Figure~\ref{fig:react-extreme-large} shows that this pattern holds for every model, not just in the average. For example, the strongest model (Ops-4.6) reduces the zero-precision rate from 56.1\% under Minimal prompting to 15.6\% under Contextual prompting and 1.1\% under Guided prompting, while its perfect-recall rate rises from 16.7\% to 47.2\% and then 66.1\%. The same qualitative transition appears for Sonnet-4 and Sonnet-4.5. This is exactly the kind of evidence that mean-only reporting would hide.

\begin{figure*}[t]
\centering
\includegraphics[width=0.97\textwidth]{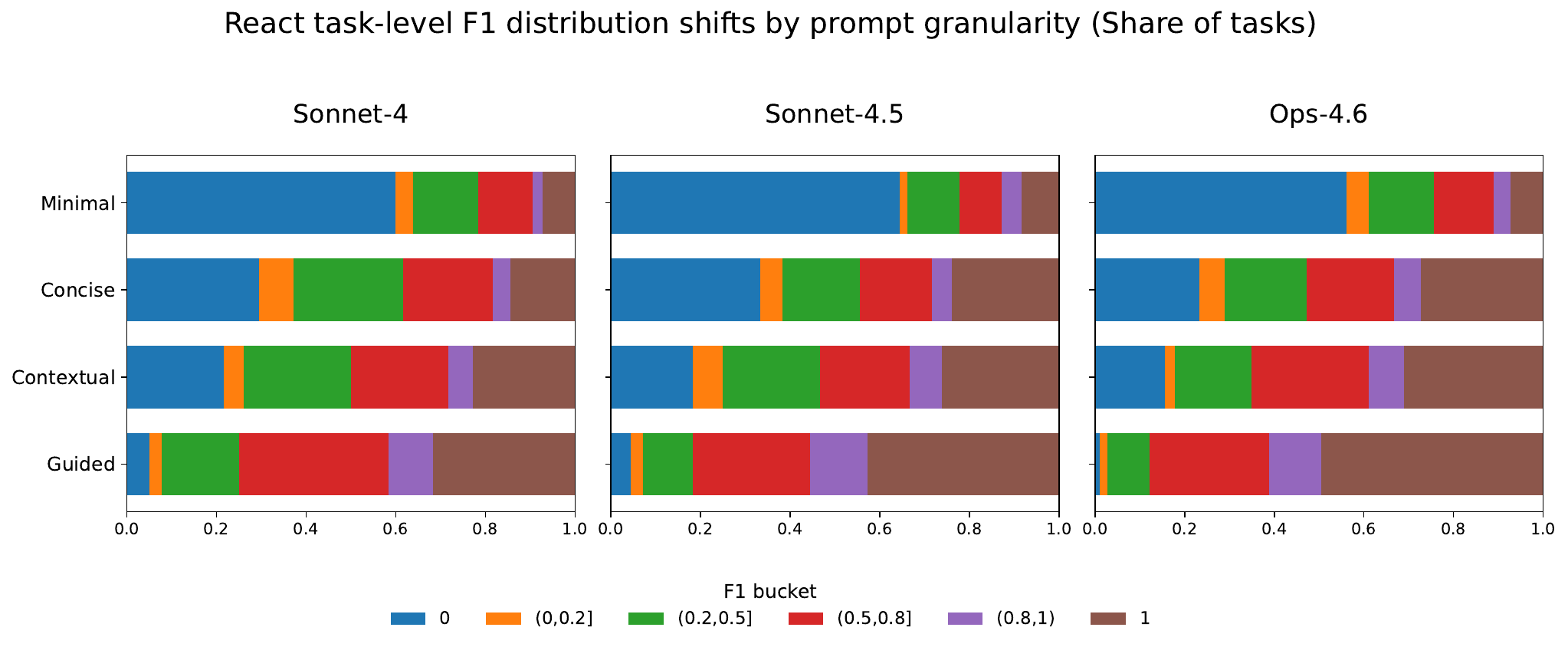}
\caption{React baseline, task-level F1 distributions. The distribution does not simply shift upward smoothly; rather, prompt strengthening moves substantial probability mass out of the zero-performance bin and into the high-F1 and exact-match bins.}
\label{fig:react-dist-large}
\end{figure*}

Figure~\ref{fig:react-dist-large} provides the complementary distributional view. The key phenomenon is a \emph{shape change}, not just a mean change. Minimal prompts produce a highly brittle regime dominated by zero-performance outcomes. Concise prompts recover some middle-performing cases, but Contextual prompts are where the benchmark begins to resemble a realistic evaluation setting: many catastrophic failures disappear, yet the task still requires repository exploration. Guided prompts continue to improve absolute performance, but they also shift a substantial amount of mass into near-perfect and perfect bins, which is precisely why they are less suitable as the default prompt setting for measuring the marginal value of repository-derived code knowledge. From a benchmark-design perspective, these figures support the claim that \textbf{Contextual prompting is the most defensible primary operating point}: it removes a large fraction of under-specification failures without saturating the benchmark as aggressively as Guided prompting.

\begin{figure*}[t]
\centering
\includegraphics[width=0.96\textwidth]{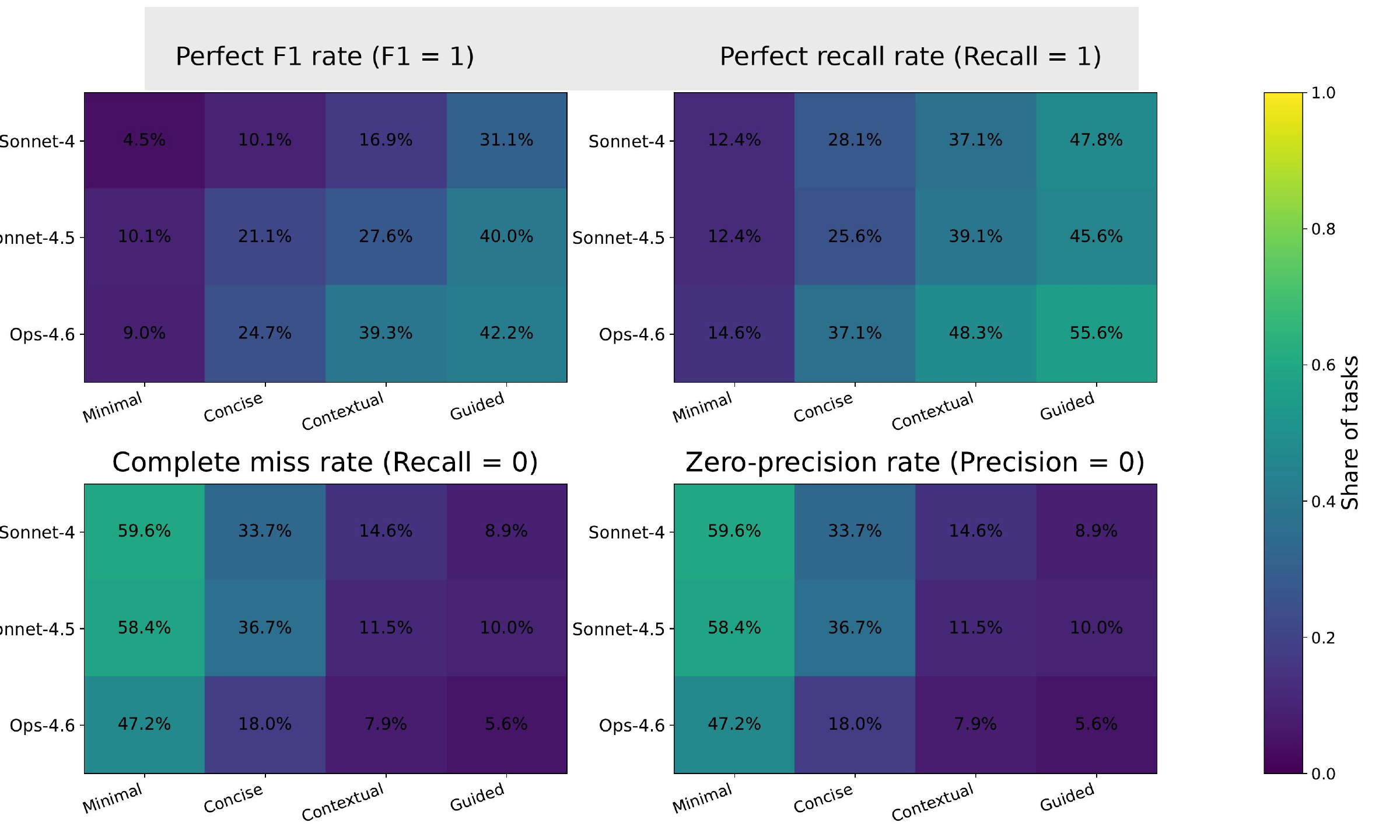}
\caption{DragonFly baseline, task-level extreme outcomes. Each cell reports the fraction of historical PR tasks that land in an exact success or exact failure state. As in React, stronger prompts sharply reduce both complete misses (Recall$=0$) and spurious localization failures (Precision$=0$), while increasing exact localization (Recall$=1$, F1$=1$). Gray cells indicate combinations for which task-level detail exports are unavailable in the current archive.}
\label{fig:dragonfly-extreme-large}
\end{figure*}

Figure~\ref{fig:dragonfly-extreme-large} shows the same qualitative transition on DragonFly. Although two Concise-condition detail exports are unavailable in the current archive, the observed cells already show the same directional behavior as React: stronger prompts reduce exact failure states and increase exact localization. This strengthens the claim that the benchmark is capturing a stable capability phenomenon rather than a repository-specific artifact.

\begin{figure*}[t]
\centering
\includegraphics[width=0.97\textwidth]{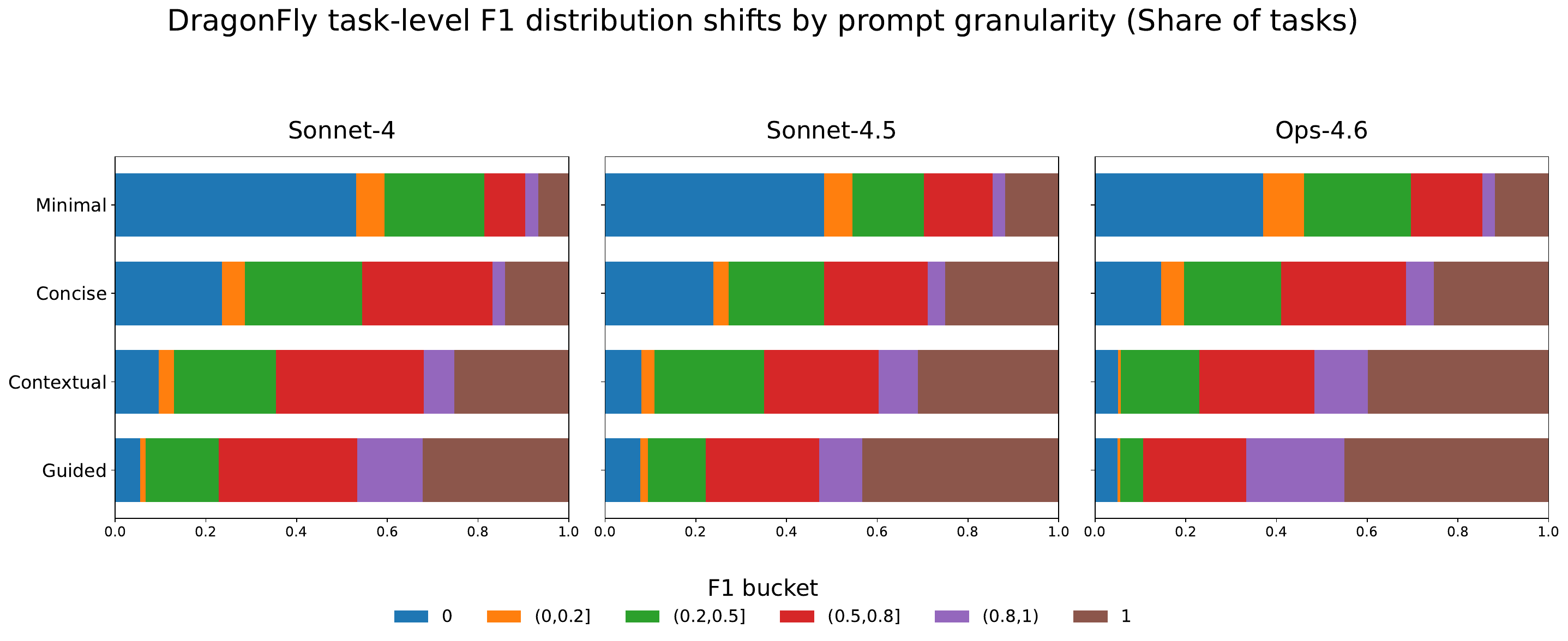}
\caption{DragonFly baseline, task-level F1 distributions. As on React, prompt strengthening moves substantial probability mass out of the zero-performance bin and into high-F1 and exact-match bins.}
\label{fig:dragonfly-dist-large}
\end{figure*}

Figure~\ref{fig:dragonfly-dist-large} provides the DragonFly distributional view. The same structural pattern appears: Minimal prompting is dominated by exact failures, Contextual prompting removes much of this brittle behavior without saturating the task, and Guided prompting shifts additional mass into near-perfect and perfect bins. This cross-repository agreement is one of the strongest pieces of evidence that the benchmark is measuring something stable.

\subsection{Cross-Repository Comparison}
\begin{figure*}[t]
    \centering
    \begin{subfigure}[t]{0.48\textwidth}
        \centering
        \includegraphics[width=\textwidth]{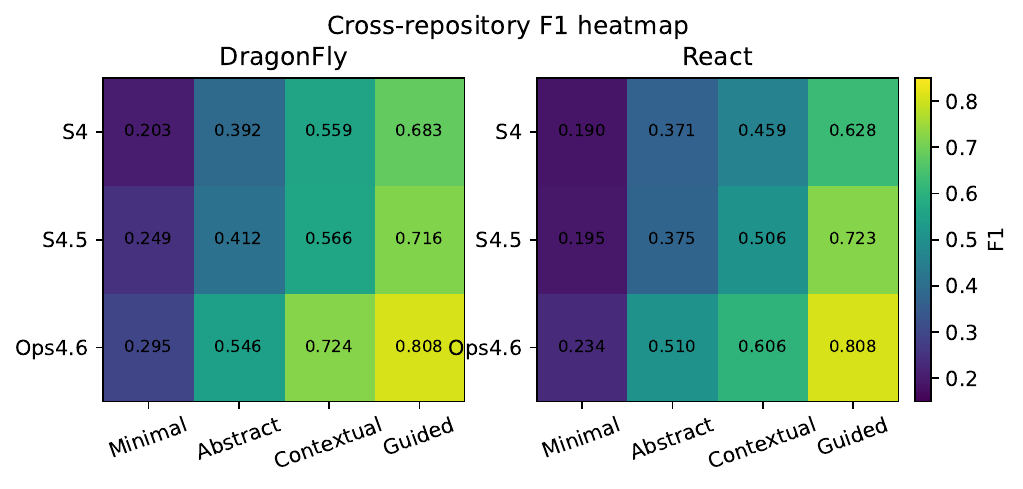}
        \caption{Cross-repository F1 heatmap.}
        \label{fig:heatmap}
    \end{subfigure}\hfill
    \begin{subfigure}[t]{0.48\textwidth}
        \centering
        \includegraphics[width=\textwidth]{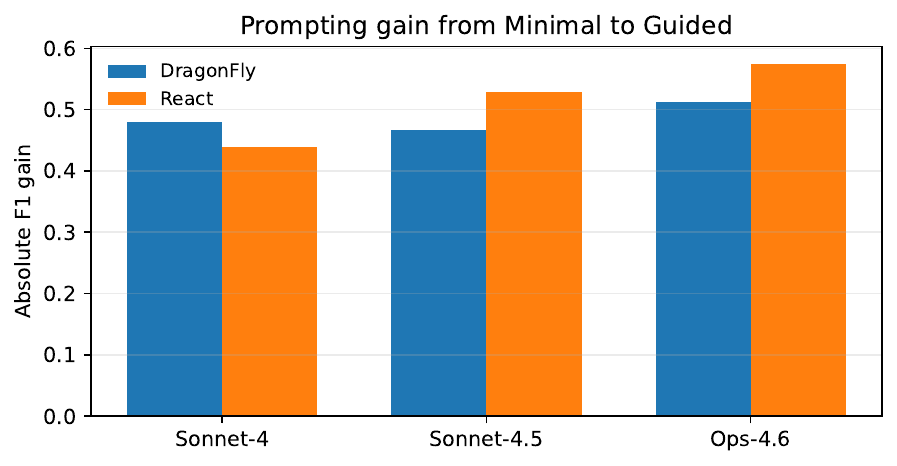}
        \caption{Absolute F1 improvement from Minimal to Guided prompting.}
        \label{fig:gain}
    \end{subfigure}
    \caption{Cross-repository comparison. Stronger prompts improve localization across both repositories, and the absolute gain from Minimal to Guided prompting is large for every model.}
    \label{fig:crossrepo}
\end{figure*}

Across both repositories, the same ordering holds almost everywhere: stronger prompts improve localization, and stronger models improve performance at each prompt level. The strongest observed scores are very similar across repositories under Guided prompting (0.8081 on DragonFly and 0.8078 on React for Ops-4.6), while weaker prompt settings expose larger repository differences. This suggests that richer task framing can partially compensate for repository difficulty, but only at the cost of reducing the need for repository exploration.

These results lead to two methodological conclusions. First, \textbf{prompt granularity must be treated as a first-class benchmark variable}. A benchmark that does not standardize how pull requests are translated into prompts will yield scores that are difficult to compare across systems. Second, \textbf{contextual prompting is the most defensible default setting} for future knowledge-augmented evaluation: it is strong enough to make tasks solvable, but still leaves room for repository-derived code knowledge to matter.

\section{Discussion}
The present experiments should be interpreted as a calibration study for the benchmark protocol rather than a full augmentation-efficacy study. Because only the baseline arm is currently reported, the results do not yet quantify the marginal gain of repository-derived code knowledge. What they do show is that benchmark outcomes can change dramatically even before repository-knowledge augmentation is introduced. Any future A/B claim must therefore be larger than, or at least clearly separable from, the variance induced by prompt construction alone.

\section{Threats to Validity and Limitations}
The paper introduces the evaluation protocol and reports baseline sensitivity analyses, but it does not yet report the paired knowledge-augmented-vs-baseline deltas that would complete the causal argument.

\textbf{Uncertainty reporting is still incomplete.} The current draft now includes task-level structural evidence, but it would be stronger with confidence intervals, explicit two-proportion tests for extreme-outcome rates, and formal distributional tests reported in a summary table.

\textbf{Prompt-generation details remain under-specified.} Although the four prompt granularities are defined conceptually (Minimal, Concise, Contextual, Guided), the paper would be stronger with exact prompt templates, the LLM conversion instructions, and explicit leakage-control filters.

\textbf{Metric scope is currently narrow.} The present experiments measure file-level localization rather than final patch correctness. File prediction is useful and often diagnostic, but it is not equivalent to solving the software engineering task end to end. A complete knowledge-augmented evaluation should therefore also report patch success, test-suite pass rate, or execution-based correctness when feasible.

\textbf{Repository coverage is still limited.} Two repositories are enough to show early cross-repository stability, but not enough to claim broad generalization.

\section{Conclusion}
We presented a time-consistent benchmark protocol for repository-knowledge-augmented systems. The protocol constructs repository knowledge at historical snapshot time $T_0$, derives evaluation tasks from pull requests merged in $(T_0,T_1]$, and compares systems under a matched A/B design that isolates repository-derived code knowledge from all other factors. We also reported a baseline characterization study across DragonFly, React, three model configurations, and four prompt granularities, showing that prompt design exerts a large and reproducible influence on repository-level file prediction. These patterns are consistent across both React and DragonFly. Despite being evaluated on two repositories, the benchmark demonstrates strong cross-repository consistency, with aligned outcome structures and model ordering. This suggests that the observed effects arise from stable properties of the evaluation protocol rather than dataset-specific artifacts.

The main implication is methodological. For repository-aware software engineering benchmarks, temporal consistency and prompt control are not auxiliary details; they are core validity requirements. More specifically, realism, freshness, and execution-based evaluation are not by themselves sufficient when the scientific question concerns the net effect of a repository-knowledge layer; matched causal control is also required. The next step is to complete the matched knowledge-augmented-vs-baseline evaluation with task-level deltas, uncertainty estimates, exact prompt templates, and broader repository coverage.

\appendix

\end{document}